\documentclass[12pt]{article}
\usepackage{ulem}
\usepackage{times}
\usepackage{color}
\usepackage{graphicx}
\usepackage{calc}
\usepackage{epsfig}
\usepackage{float}
\usepackage{amsfonts}
\newfloat{Continuation}{htb}{lob}
 \fussy \oddsidemargin0.0cm \evensidemargin0.0cm
\begin{document}

\parskip=0.2in
\baselineskip=0.3in

\noindent{\large\bf Title}

\noindent The information transmitted by spike patterns in single
neurons.

\noindent{\large\bf Authors}

\noindent Hugo G. Eyherabide and In\'es Samengo

\noindent{\large\bf Affiliations}

\noindent Centro At\'omico Bariloche and Instituto Balseiro,
R8402AGP, San Carlos de Bariloche, Argentina.

\noindent{\large\bf Corresponding Author}

\noindent In\'es Samengo, Centro At\'omico Bariloche, R8402AGP,
San Carlos de Bariloche, R\'{\i}o Negro, Argentina. Tel: ++ 54
2944 445100 (int: 5391). Fax: ++54 2944 445299. \\
\noindent Email: samengo@cab.cnea.gov.ar

\noindent{\large\bf Abstract}

\noindent Spike patterns have been reported to encode sensory
information in several brain areas. Here we assess the role of
specific patterns in the neural code, by comparing the amount of
information transmitted with different choices of the readout neural
alphabet. This allows us to rank several alternative alphabets
depending on the amount of information that can be extracted from
them. One can thereby identify the specific patterns that constitute
the most prominent ingredients of the code. We finally discuss the
interplay of categorical and temporal information in the amount of
synergy or redundancy in the neural code.

\noindent{\large\bf Keywords}

\noindent neural code, sensory encoding, information theory, spike
patterns, synergy.

\pagebreak

\section{The role of spike patterns in the neural code}
\label{s1}

Many studies have shown that precise spike timing plays an important
role in the encoding of sensory stimuli. For example, in the cat LGN
(Reinagel and Reid, 2000) the transmitted information was shown to
be a rapidly increasing function of the resolution with which the
timing of spikes was read out, up to the sub-millisecond regime.
Similar results were observed in the rat somatosensory cortex
(Panzeri et al., 2001; Arabzadeh et al., 2006), the fly H1 neuron
(Strong et al., 1998), and the grasshopper auditory system (Rokem et
al., 2006).

In some cases, the additional information obtained by reading out
the neural responses with high temporal resolution was explained in
terms of a firing-rate code whose time-dependent spiking probability
exhibited sharp and rapid temporal fluctuations accurately encoded
by just a few spikes (Montemurro et al., 2007). In such cases, the
neural code is mainly instrumented by precisely timed single spikes.
In other cases, however, extended spike patterns (comprising more
than a single action potential) have been found to play an important
role. This was shown, for example, in the cat LGN (Reinagel and
Reid, 2000), and in the cat auditory cortex (Furukawa and
Mindelbrooks, 2002), by comparing the information transmitted by the
original spike train with that encoded in shuffled responses, where
the within-trial correlations between subsequent spikes were
eliminated, though preserving the timing precision of individual
spikes. These studies concluded that specific temporal arrangements
of spikes were relevant to information transmission.

Several distinct features within spike trains have been identified
as possible carriers of the neural message. For example, both in the
primate primary visual cortex (Reich et al., 2000) and in the
electrosensory lobe of the weakly electric fish (Oswald et al.,
2007) the duration of inter-spike intervals (ISIs) was shown to
encode particular stimulus properties, i.e., ISIs of different
lengths were associated to distinct stimulus features. In the
salamander retina (Gollisch and Meister, 2008), a prominent coding
role was assigned to the correlations between the response latencies
of different cells. In the leech, the velocity of tactile stimuli
was encoded by the number of spikes in each burst (Arganda et al.,
2007).

The information that is encoded in structured sequences of spikes
and silences is only available when the spike train is read out
with extended words, that is, with temporal windows that are
sufficiently long to contain several spikes. To quantify the extra
information that is gained from such long read-out windows, the
concepts of {\sl synergy} and {\sl redundancy} have been
introduced. In general terms, if the information carried by a
collection of variables is higher than the sum of the information
carried by the individual elements, then the code is called
synergistic. In the opposite situation, the code is redundant
(Brenner et al., 2000; Pola et al., 2003; Schneidman et al.,
2003). This means that, when assessing the amount of synergy
between two variables $R_1$ and $R_2$, we need to determine
whether the information $I(R_1, R_2; S)$ encoded by the joint pair
$(R_1, R_2)$ about another variable $S$ is larger or smaller than
the sum $I(R_1; S) + I(R_2; S)$ of the information encoded by
$R_1$ and $R_2$ separately. The amount of synergy $S_{\rm syn}$ is
(Pola, 2003; Schneidman 2003)
\begin{equation}
S_{\rm syn} = I(R_1, R_2; S) - I(R_1; S) - I(R_2; S). \label{e6}
\end{equation}

When applying these ideas to neural coding, $S$ is typically the
stimulus, and $R_1$ and $R_2$ are two features of the neural code.
In population coding, for example, $R_1$ and $R_2$ are the responses
of two different neurons (Gawne and Richmond, 1993; Petersen et al.,
2001). In this paper we focus on single-neuron coding, so $R_1$ and
$R_2$ represent the response of just one neuron in different time
bins.

The amount of synergy between time bins in the neural code has
been often estimated in order to determine whether spike patterns
play a relevant role (Reinagel and Reid, 2000; Liu et al., 2001;
Kumbhani et al., 2007). However, even when $S_{syn}$ was found
positive, the information-bearing spike patterns were not
explicitly displayed. In this paper we focus on specific spike
patterns, and we provide a quantitative assessment of their
relevance in the neural code. We do so by calculating the mutual
information between stimuli and responses for different choices of
the read-out alphabet. Each alphabet is composed of a collection
of spike patterns. As we vary the alphabet, we make increasingly
finer distinctions between patterns. By assessing how much
information is gained by making fine discrimination between
patterns, we reveal which are the informative patterns. Finally,
we discuss the consequences of our analysis in the evaluation of
how synergistic or redundant the neural code is.

For definiteness, we focus on bursting neurons, where the
intra-burst spike count $n$ (that is, the number of spikes within
each burst) is taken as the relevant response feature encoding
specific stimulus properties. Burst firing has been ubiquitously
found in sensory systems (see, for example, Krahe and Gabbiani,
2004; and references therein). There are different dynamical
processes that produce bursting. For example, in the mammal LGN
(Sherman, 2001) burst generation is associated to a T-type calcium
current, that induces slow oscillations in the membrane potential
(low threshold spikes) which, in their depolarized phase, give rise
to burst firing. In the weak electric fish, instead, bursting
results as a consequence of the geometry of cells: the active
propagation of action potential into the dendrites produces a
rebound effect, that induces high-frequency repetitive firing
(Mainen and Sejnowsky, 1996). In the grasshopper auditory system
(Eyherabide et al., 2008) bursting is induced by specific
time-dependent stimuli, that fluctuate with time scales of the order
of 5-10 msec.

Several studies have demonstrated that bursts have a specific role
in the encoding of sensory information. Thus, by comparing the
average stimulus eliciting bursts with the one generating isolated
spikes, cat LGN bursts were shown to appear in response to
stimulus features having a significantly lower power spectrum
(Lesica and Stanley, 2004), higher contrast (Reinagel et al.,
1999; Lesica and Stanley, 2004) and comprising more natural
statistical distributions (Lesica and Stanley, 2004; Denning and
Reinagel, 2005) than the features associated with the generation
of isolated spikes. Similarly, in the electric fish, bursting was
associated with particularly low-frequency events (Oswald et al.,
2004), comprising either excitatory or inhibitory stimulus
deflections (Metzner et al., 1998). Moreover, the detailed
properties of the burst-triggering features were shown to depend
critically on the mean stimulus level (Lesica et al., 2006), thus
suggesting an adaptive, dynamical role in burst coding.

In these studies, the stimulus features encoded by the event {\em
burst} (as a whole) were found significantly different by those
encoded by the event {\em isolated spike}. Other analyses have
gone further, discriminating the features encoded by bursts
containing specifically $n$ spikes. If different $n$ values are
associated to different stimulus features, then we may speak of a
non-trivial neural code, where the intra-burst spike count $n$
carries information about the stimulus. So far, $n$ was shown to
encode the magnitude of the slope of the stimulus in bursting
pyramidal neurons (Kepecs et al., 2002; Kepecs and Lisman, 2003),
the velocity of tactile stimuli in leech (Arganda et al., 2007),
and the amplitude of acoustic fluctuations in grasshopper auditory
neurons (Eyherabide et al., 2008).

\section{Methods}

\subsection{Estimation of mutual information rates}
\label{s3p1}

\begin{figure}[htdf]
\centerline{\includegraphics{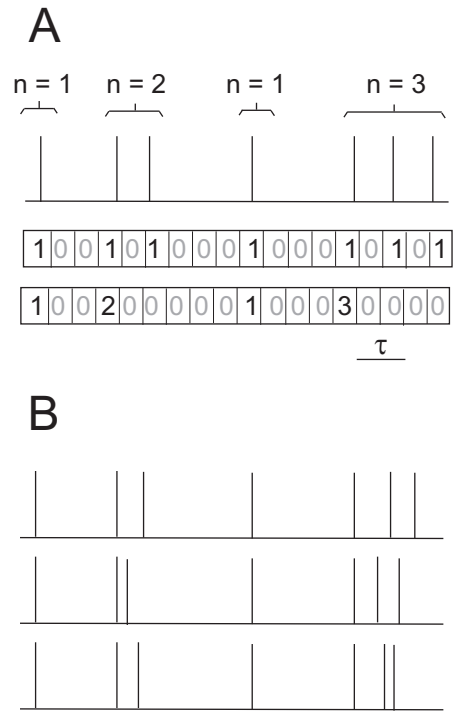}} \caption{\label{fig1}
Representation of the spike train as a sequence of symbols. {\em
A}: A given spike train (top) is represented as a binary string
(top string) or as an integer string (bottom string), depending on
the alphabet used to identify response patterns. In the integer
string, all consecutive spikes separated by ISIs smaller than
$\tau$ were grouped into the same burst. {\em B}: Three example
spike trains that are mapped onto the same integer string, though
corresponding to different binary strings. }
\end{figure}

In order to calculate mutual information rates between stimuli and
neural responses, we have adapted the {\em Direct Method}
introduced by Strong et al. (1998) to estimate the information
transmitted by specific spike patterns. In its original
formulation, the method begins by representing the spike train as
a sequence of integer numbers, where each number indicated the
number of spikes falling within the corresponding time bin (see
Fig.~\ref{fig1}).
For high-resolution binnings, spike trains are always represented as
binary strings, since the duration of the spike itself ($\approx 1$
msec) and of the refractory period ($\approx 2-3$ msec) forestalls
the occurrence of more than a single spike in a short time bin.

Here, the method was broaden to also encompass estimations of the
information carried by specific spike patterns. In the first
place, one must decide which are the spike patterns that are going
to compose the building-blocks of the neural alphabet. For
definiteness, in the top string of Fig.~\ref{fig1}{\em A}, we show
the representation of a given spike train, when the only symbols
comprising the alphabet are spike (1) and no spike (0). In the
bottom string, however, we use a different alphabet. In this case,
each time bin represents the number of spikes that follow, whose
inter-spike intervals (ISIs) are smaller than a certain prescribed
value (equal to $\tau$, in the case of the figure). Hence, the
neural alphabet is composed of all non-negative integers, and each
symbol represents the intra-burst spike count $n$ of the burst
that begins at that time bin.

Notice that once we have decided which spike patterns are going to
be taken into account, the new representation of the spike train
in terms of patterns can be derived from the original binary
representation. As such, it constitutes a processed version of the
spike train. The {\em Data Processing Theorem} (Cover and Thomas,
1991) precludes any manipulation of the spike train from
increasing the information transmitted about the stimulus. Hence,
the information encoded by the selected patterns is necessarily
smaller than the originally contained in the spike train. In fact,
there might be several binary sequences that are mapped onto the
same integer sequence. In our example, where bursts are defined as
sequences of spikes whose ISIs fall below $\tau$, the number $n$
is allocated at the time bin where the burst begins. Hence, with
this representation, the information about the precise temporal
location of the subsequent spikes in the burst is lost. In
Fig.~\ref{fig1}{\em B}, three example spike trains corresponding
to the same integer string in {\em A} are exhibited.

Once the spike train is represented as a sequence of symbols (one
symbol per pattern), the method follows the same steps as in
Strong et al. (1998). That is, consecutive symbols are grouped
into {\em words} of length $w$, and the noise entropy of the
distribution of words is subtracted from the total entropy. One
thus obtains the information $I_w$ carried by words of length $w$.
The analysis is repeated for different word lengths $w$, and the
mutual information rate is
\begin{equation}
I = \lim_{w \to \infty} \frac{I_w}{w}. \label{e1}
\end{equation}
In this paper, in order to assess which patterns are relevant to
information transmission, we have calculated the information rate
$I$ with different choices of the patterns conforming the neural
code.

When estimating the information transmitted by the full unprocessed
spike train, we have worked with time bins of 0.1 msec (for the
simulated data) and 0.4 msec (for the experimental data). These
numbers were selected small enough to capture the temporal precision
of the responses, and at the same time, not too small as to yield
severe sampling problems. Bias corrections were taken into account
using the NSB method (Nemenman et al., 2004).

By taking the limit to infinite word length in Eq.~(\ref{e1}), we
are considering arbitrarily long words. In practice, the limit
only needs to be taken until the window is long enough as to
encompass all the correlation structure in the spike train. Even
larger windows need not be considered, since once subsequent
windows are independent from each another, $I_w$ grows linearly
with $w$, so the ratio in Eq.~(\ref{e1}) remains constant. When
$w$ is smaller than the correlation length in the spike train,
however, $I_w/w$ may bear a non-trivial dependence with $w$.

\subsection{Simulated data}
\label{s3p2}

In the simulations, the stimulus consists of a Poisson command
signal, that has four different aspects: $\alpha, \beta, \gamma$ or
$\delta$, as shown in Fig.~\ref{fig2}. For definiteness, we can
think of the stimulus as a sequence of brief light pulses, that can
appear in four different colours. When the red light flashes, the
stimulus is labelled as $\alpha$. In the same way, blue is
associated with $\beta$, green and yellow with $\gamma$ and
$\delta$. Each stimulus induces a response burst of $n$ spikes, and
the value of $n$ depends on the identity of the stimulus. The
temporal axis is segmented into bins of 0.1 msec, and the total
duration of the stimulus is 100 sec. If $P_0$ represents the
probability that no stimulus was shown, the stimulus probabilities
in each time bin are
\begin{equation}
\begin{array}{ll}
P_0 = 0.995314 & \\
P_\alpha = 1.7378 \ 10^{-3} & P_\beta = 1.2874 \ 10^{-3} \\
P_\gamma = 9.5373 \ 10^{-4} & P_\delta = 7.0654 \ 10^{-4}
\end{array}
\label{e1a}
\end{equation}
These probabilities were chosen so that the four stimuli were
ranked in order of decreasing importance. If due to limited
resources a channel cannot afford to represent the identity of all
the four stimuli, their relative probabilities make it convenient
to discriminate first stimulus $\alpha$ from the other three, then
(if more resources are available) stimulus $\beta$ from the
remaining two, and finally (if resources are enough), stimuli
$\gamma$ and $\delta$. This scenario implies there is a hierarchy
of possible codes, depending on the length of the code-words that
we are willing to use.

The entropy rate of the stimulus is
\begin{equation}
H_{\rm stim} = - \frac{1}{\Delta t}\sum_i P_i \ \log(P_i),
\end{equation}
where $i$ runs over $0, \alpha, \beta, \gamma, \delta$, and $\Delta
t$ is the bin size. This entropy can be decomposed into temporal
entropy and categorical entropy,
\begin{equation}
H_{\rm stim} = H_{\rm temporal} + H_{\rm categorical}.
\label{edescompsition}
\end{equation}
The temporal entropy is the one that only distinguishes whether at
each point in time there is a stimulus or not. Hence, if $P_1 =
P_\alpha + P_\beta + P_\gamma + P_\delta$ represents the
probability that any stimulus is shown,
\begin{equation}
H_{\rm temporal} = - \frac{1}{\Delta t} \left[ P_0 \log (P_0) + P_1
\log(P_1) \right].
\end{equation}
The categorical entropy quantifies the variability associated to the
fact that each stimulus can appear in four different types
\begin{equation}
H_{\rm categorical} = \frac{P_1}{\Delta t} \left[- \sum_{i =
\alpha,\beta,\gamma,\delta} \frac{P_i}{P_1} \log
\left(\frac{P_i}{P_1} \right)\right]. \label{ecategorical}
\end{equation}
In Eq.~(\ref{ecategorical}), the term between brackets is formally
an entropy, with normalized probabilities $P_i/P_1$.

In our work, the true entropy rate of the stimulus is compared to
the one of another random process, where only a single type of
stimulus is shown, and whose apparition rate is equal to $P' =
P_\alpha + 2 P_\beta + 3 P_\gamma + 4 P_\delta$.

\subsection{Electrophysiology}
\label{s3p3}

Intracellular recordings were conducted {\sl in vivo} on the
auditory nerve of {\em Locusta Migratoria} (see the details in
Rokem et al., 2006). The auditory stimulus was a high frequency
sine tone (3 kHz) modulated with a low-pass filtered Gaussian
amplitude distribution of controlled mean (44.76 dB), standard
deviation (6 dB) and cut-off frequency (200 Hz). This stimulus was
designed to incorporate some key features of grasshopper courtship
songs, are accurately represented in the spike trains (Rokem et
al., 2006) and are relevant to behaviour (Balakrishnan et al.,
2001; Krahe et al., 2002). The acoustic stimulus lasted for 1
second, and was repeated 503 times, while the spike train was
recorded. Between successive repetitions, a pause of 700 msec was
incorporated, in order to avoid slow adaptation effects. Bursting
responses were observed, and the probability of observing a burst
of exactly $n$ spikes in a bin of 0.4 msec was
\begin{equation}
\begin{array}{ll}
P({\rm no \ event}) = 0.96 \\
P(n = 1) = 2.89 \ 10^{-2} & P(n = 2) = 6.71 \ 10^{-3} \\
P(n = 3) = 1.42 \ 10^{-3} & P(n = 4) = 2.52 \ 10^{-4} \\
P(n = 5) = 4.36 \ 10^{-5}& P(n = 6) = 2.4 \ 10^{-6}
\end{array}
\end{equation}

\section{Quantifying the information transmitted by specific spike
patterns: a model study}
\label{s4}

\begin{figure}[htdf]
\centerline{\includegraphics{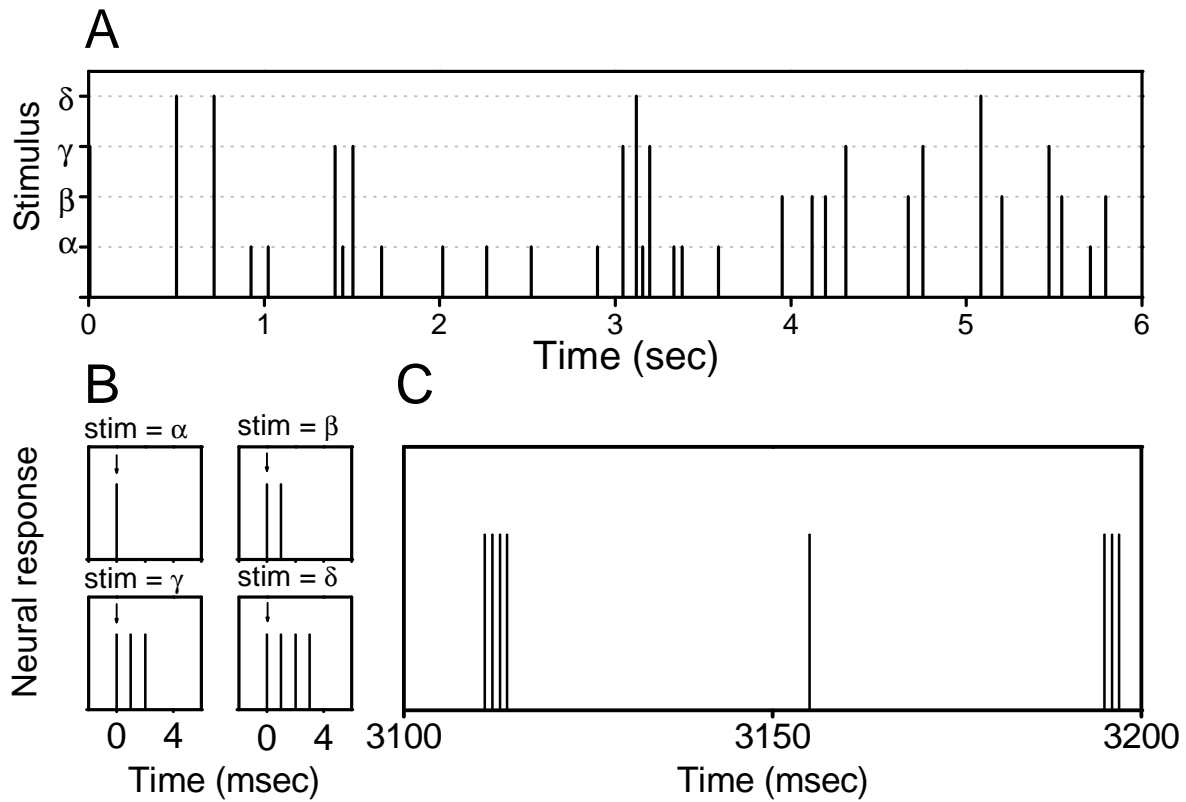}} \caption{\label{fig2}
Sketch of a typical stimulus stretch, and the associated
responses. {\em A}: The stimulus is a Poisson process, with 4
possible different outcomes (here represented by the height of the
vertical bars). {\em B}: Neural responses associated with each of
the four stimuli. Different stimuli generate bursts of different
spike count $n$. {\em C}: Temporal evolution of the responses
corresponding to a small interval of the stimulus in {\em A}.
Notice that {\em A} is depicted in seconds, whereas {\em B} and
{\em C} are scaled in milliseconds.}
\end{figure}

In the simulations, the stimulus consisted of a Poisson command
signal, that could take the values $\alpha, \beta, \gamma$ or
$\delta$, as shown in Fig.~\ref{fig2}. The four possible outcomes
should be interpreted as four
particular stimuli (or stimulus features) that pop out
occasionally, and induce four different responses in a sensory
cell. At any given time bin, either there is no stimulus (the null
event), or stimulus $\alpha$, $\beta$, $\gamma$ or $\delta$ is
presented to the subject. Assigning the label 0 to the null event,
we are in the presence of a process with 5 possible outcomes, with
probabilities $P_0, P_{\alpha}, P_{\beta}, P_{\gamma}$ and
$P_{\delta}$. Subsequent time bins are independent from one
another. In Fig.~\ref{fig2}{\em A}, an example stimulus is
depicted. The height of the vertical bars represents which of the
four stimuli was shown.

We assume that the neuron encodes the identity of the different
stimuli with bursts of varying spike count $n$, modelling the
experimental results discussed above (Kepecs et al., 2002; Kepecs
and Lisman, 2003; Arganda et al., 2007; Eyherabide et al., 2008).
As illustrated in Fig.~\ref{fig2}{\em B}, the cell generates a
single spike in response to stimulus $\alpha$, a doublet in
response to stimulus $\beta$, a triplet for stimulus $\gamma$ and
a quadruplet for $\delta$. Notice that, by construction, the
identity of the stimulus is entirely encoded in the intra-burst
spike count $n$. Hence, in order to be able to discern which
stimulus was presented, a downstream neuron (or an external
observer, as ourselves) should read out the activity of the cell
using extended words. This neural code has two distinctive
components. First, the time at which a burst is initiated encodes
{\sl when} a stimulus was presented. We then say that the spike
train carries {\sl temporal} information. Second, the number of
spikes inside the burst encodes {\sl which} stimulus was presented
(Theunissen and Miller, 1995), that is, provides {\sl categorical}
information about the stimulus. As shown below, these two aspects
are sometimes confused, if unappropriate readouts are employed.

For simplicity, we assume that the input/output transformation is
noiseless. Or, equivalently, that spike-time jitter is always
inside the size of each bin, that is, 0.1 msec. Hence, the
conditional response probabilities read
\begin{equation}
\begin{array}{ll}
P({\rm burst \ of \ order \ }n|\alpha) = \delta^k(n, 1) & P({\rm
burst \ of \ order \ }n|\beta) = \delta^k(n, 2) \\
P({\rm burst \ of \ order \ }n|\gamma) = \delta^k(n, 3) & P({\rm
burst \ of \ order \ }n|(\delta) = \delta^k(n, 4)
\end{array}
\label{e2}
\end{equation}
where the Kronecker Delta function $\delta^k(n, i) = 1$ if $n = i$,
and vanishes otherwise. Our theoretical neuron, therefore, is not
defined through a stochastic or a dynamical process, but rather
operates as a mere transcription device, that represents stimuli in
terms of spikes. In this paper, we focus on the question of how
faithfully this translation can be read out, depending on the
readout alphabet and word length we use.

For any coding scheme, the information $I_w$ contained in words of
length $w$ must always be an increasing function of $w$, simply
because the longer the words we are reading out, the more we know
about the stimulus. If time bins are fully independent from each
other, then $I_w$ grows linearly with $w$ (Cover and Thomas,
1991). If time bins are correlated, then $I_w$ may grow faster or
slower than linearly, thus giving rise to interesting
dependencies. In these cases, $I_w/w$ may either be an increasing
or a decreasing function of $w$, for intermediate $w$. If the
encoding system, however, has a finite correlation time, then
eventually, for sufficiently long word lengths, $I_w$ will grow
linearly with $w$. There, $w$ can be assumed to contain all the
relevant code-words used to transmit information, and $I = I_w/w$
can be safely considered the {\sl real} information rate.

\begin{figure}[htdf]
\centerline{\includegraphics[keepaspectratio=true, clip = true,
scale = 0.95, angle = 0]{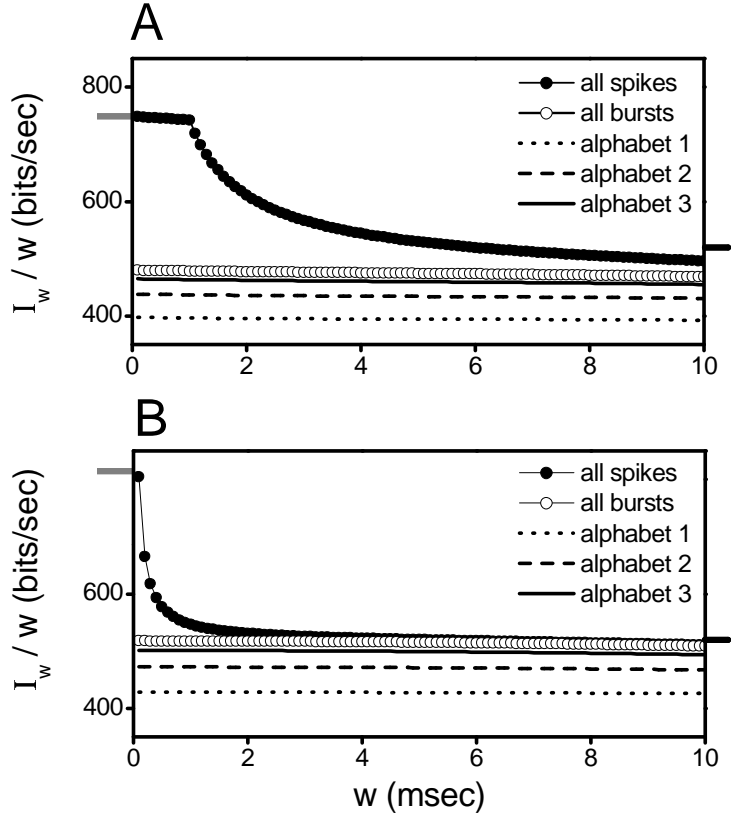}} \caption{\label{fig3}
Information ratio $I_w / w$ as a function of the word length $w$.
Different lines and symbols correspond to different
identifications of the neural alphabet. The information
transmitted by the whole collection of spikes shows a dependence
with the word-length, tending asymptotically to the information
encoded by the whole collection of bursts. {\em A}: Refractory
period: 1 msec. {\em B}: Refractory period: 0.1 msec. Grey
horizontal segments on the left: analytically-calculated entropy
rate of a binary Poisson stimulus, whose mean frequency is equal
to the firing rate of the model neurons in {\em A} and {\em B}.
Black horizontal segments on the right: analytically-calculated
entropy rate of the stimulus.}
\end{figure}

In Fig.~\ref{fig3}, the rate $I_w / w$ is depicted as a function of
the word length $w$. The desired information rate $I$ is defined
as the limit of this ratio when $w \to \infty$ (see Methods,
Sect.~\ref{s3p1}). Black circles represent the information
transmitted by the whole collection of spikes with the usual
binary code (1 = spike, 0 = no spike). White circles and lines
represent other codes (see below). The two panels correspond to
two different choices of the refractory period $\tau$ of the
intra-burst ISI. In {\em A}, $\tau$ is equal to 1 msec, as in
Fig.~\ref{fig2}{\em B}, whereas in {\em B}, $\tau = 0.1$ msec.

Noiseless conditional probabilities imply that the noise entropy
of the responses vanishes. Hence, in this case, the mutual
information is equal to the total entropy of the responses. This
entropy, in turn, should be expected to lie slightly below the
entropy of the stimulus, which can be evaluated analytically (see
Methods, Sect.~\ref{s3p2} for the relevant parameters). In
Fig.~\ref{fig3}{\em A}, we see that in the limit of long words,
$I_w/w$ approaches the theoretical value of the stimulus entropy
(represented by the black bar bulging outwards, at the right of
the plot), though falling slightly below. This small discrepancy
is due to the fact that even noiseless spike trains cannot always
represent the stimulus faithfully. If, for example, two
consecutive stimuli $\delta$ are drawn in two sequential time
bins, the system has no time to allocate four spikes for each
stimulus. Recall that each burst of 4 spikes lasts for $4 \tau =
4$ msec, and each time bin lasts 0.1 msec. In our simulations,
hence, if a stimulus arrives before the system has finished
representing the previous stimulus, the new stimulus is ignored.
In the case of Fig.~\ref{fig3}{\em A}, this happens 420 times, out
of 4689 stimulus presentations. As $\tau$ diminishes, the overlap
probability decreases. In panel {\em B}, the results of another
simulation are depicted, where the refractory period $\tau$ was
set to 0.1 msec. This is not a biologically realistic value, so
this case should be taken as an academic exercise designed to show
that whenever the number of overlaps decreases (it has now dropped
to 21/4775), the asymptotic value of $I_w/w$ approaches the
theoretical value (the dark bar on the right) more closely.

Let us now focus on the dependence of the information with the
word length. In Fig.~\ref{fig3}{\em A}, we see that the
information encoded by the whole collection of spikes is initially
high. Why does this value surpass the theoretical stimulus entropy
rate in about 40\%? For short enough words, there is at most a
single spike in each word, and spikes appear to be positioned
anywhere within the word. For longer words, this apparent freedom
is no longer seen, since the correlations between spikes limit the
number of ways in which spikes can be located. But this limitation
is not evident, for $w < \tau$. Given that in this example there
is no trial-to-trial variability, all the apparent diversity in
the location of the spikes has to be assigned to the stimulus: all
spiking times can be assumed to be encoding something about the
stimulus. In the present case, where $w$ is too short to contain
complex spike patterns, each spike is taken to represent a
detected stimulus - even the second, third and fourth spikes of
each burst. Hence, by reading the neural activity with too short
windows, what in reality is categorical information about the
identity of the stimulus is interpreted as temporal information.
Of course, single spikes can only encode a single type of
stimulus, because a spike standing alone cannot discriminate
between different kinds of stimuli. Thus, while the categorical
information in the actual stimulus is lost, additional temporal
information about another {\em non-existing} stimulus of higher
rate is gained. In fact, the grey bars protruding outwards at the
left of the plots represent the theoretical entropy rate of a
binary Poisson process whose mean frequency is equal to the firing
rate of the modelled cells in {\em A} and {\em B} (See Methods,
Sect.~\ref{s3p2}). Notice the difference: the real stimulus has
four possible categories ($\alpha, \beta, \gamma, \delta$), and an
apparition rate of 46.85 Hz. The apparent stimulus, instead, has
only a single category, and an apparition rate of 91.05 Hz. The
tight match between the entropy of the apparent stimulus and the
value of $I_w/w$ for $w < \tau$ cannot be casual. It confirms that
for small time windows, the information rate appears larger than
the real information, because the readout system is interpreting
each spike as a new stimulus, whose frequency of occurrence is
higher than the frequency of the real stimuli. In the present
example, there is no trial-to-trial variability (or equivalently,
spike-time jitter is constrained to be smaller than our time bin
of 0.1 msec). Consequently, the additional temporal information
about the fictitious Poisson binary stimulus is significantly
high. In contrast, the lost categorical information about the real
stimulus is limited. The tradeoff between these two effects
explains the difference between the initial and final value of
$I_w/w$.

However, as soon as the window length $w$ reaches $\tau$ the
information rate diminishes rapidly. This happens at $w = 1$ msec,
in Fig.~\ref{fig3}{\em A}, and at $w = 0.1$ msec, in {\em B}. At
this point, two spikes may fall inside the same window, so the
apparent freedom with which spikes had seemed to be located is no
longer present: often, spikes come in close succession, separated by
an interval $\tau$. There is, hence, a typical correlation time in
the spike train, that reduces the number of possible words that
appear. In this example, it turns out that such correlations are the
crucial (and the unique) aspect of the spike train encoding the
identity of the stimulus. Therefore, if the information rate
diminishes as a function of $w$, it is not due to the fact that
correlations provide redundant information, but rather, that for
very short windows, the information was erroneously high.

What happens if we represent the spike train as a sequence of
bursts, instead of a sequence of spikes? If inside the bin $t$ we
indicate the number of spikes $n$ of the burst starting at $t$, the
ratio $I_w/w$ does not depend on the word length $w$, as shown by
the white dots in Fig.~\ref{fig3}. Therefore, the information rate
can be estimated equally well with $w = 1$ and $w \to \infty$. By
construction, in this case an $n$-based code captures all the
information contained in the spike train, and does so with 1-bin
words. This example serves to emphasize that if one is able to
identify the relevant patterns in the spike train, employing symbols
that represent those patterns explicitly can save us an enormous
amount of time and resources. Recall that the computational time
required to estimate information rates with the Direct Method grows
exponentially with $w$.

The importance of identifying the relevant patterns in the spike
train, however, goes beyond an operational convenience. By trying
out different possible encoding schemes and comparing the resulting
information rates, one may actually deduce which symbols are the
essential ingredients of the neural code. A code based on these
symbols, thus, constitutes the minimal code preserving most of the
information, and still allows us to quantifying the categorical and
the temporal information separately. In Fig.~\ref{fig3}, this is
shown by the lines. The dotted line ({\em Alphabet 1}) corresponds
to a representation of the spike train where only the time at which
a burst (any burst) is initiated. Thus, here we also use a binary
string, but now each symbol ``1'' tags the time at which a burst
with any number of spikes (including bursts of 1 spike, that is,
isolated spikes) is generated. With such a read-out code we are not
discriminating between the four different types of stimuli. All the
categorical information is lost, whereas all the temporal
information is preserved. By comparing the black circles with the
dotted line in Fig.~\ref{fig3}{\em A}, we see that there is a loss
of information in approximately 16\% ($\approx$ 76 bits/sec). In
this stimulus, hence, 84\% of the information corresponds to
determining the temporal location of each stimulus (a task for which
alphabet 1 is well suited), and 16\% to identifying which of the
four stimuli was presented (something this code fails to do
altogether).

We can go one step further, and distinguish one response pattern
from the other three. This is equivalent to assuming that isolated
spikes ($n$ = 1) encode certain stimuli, and bursts ($n > 1$)
encode some other stimuli, irrespective of whether $n = 2, 3$ or
4. If the stimuli associated with isolated spikes have
significantly more behavioural relevance than the other stimuli,
or if they appears more often, then discriminating these stimuli
from the rest may be a convenient (and economic) strategy. As
mentioned in Sect.~\ref{s1}, several experimental studies have
shown examples where bursts (as a whole) encode different stimulus
features than isolated spikes. In our example, hence, we define
{\em Alphabet 2} as composed of 3 symbols: ``0'' representing no
spike, ``1'' for isolated spikes, and ``2'' for doublets, triplets
and quadruplets of spikes. With this representation, we obtain the
information rates depicted in dashed lines. Once again, $I_w/w$ is
independent of $w$, and its numerical value accounts for 50\% of
all the categorical information in Fig.~\ref{fig3}{\em A}. This
means that out of the 76 bits/sec that are needed to identify
which stimulus was presented, 38 of them correspond to identifying
stimulus $\alpha$ from the other three. Defining {\em Alphabet 3}
as the one distinguishing between no spikes, isolated spikes,
doublets, and bursts containing either 3 or 4 spikes
indifferently, we get the information rates shown with the solid
line. Hence, the distinction between stimulus $\beta$ from
$\gamma$ + $\delta$ takes 25 extra bits/sec. Finally, the cost of
distinguishing between $\gamma$ and $\delta$ is 13 bits/sec.

The information rates obtained for read-out codes that are
increasingly complex, hence, provide a natural way of quantifying
how much information is encoded by each pattern, and how much each
pattern adds to the total information rate.

\section{Quantifying the information transmitted by specific spike patterns: an experimental study}
\label{s5}

\begin{figure}[htdf]
\centerline{\includegraphics[keepaspectratio=true, clip = true,
scale = 0.85, angle = 0]{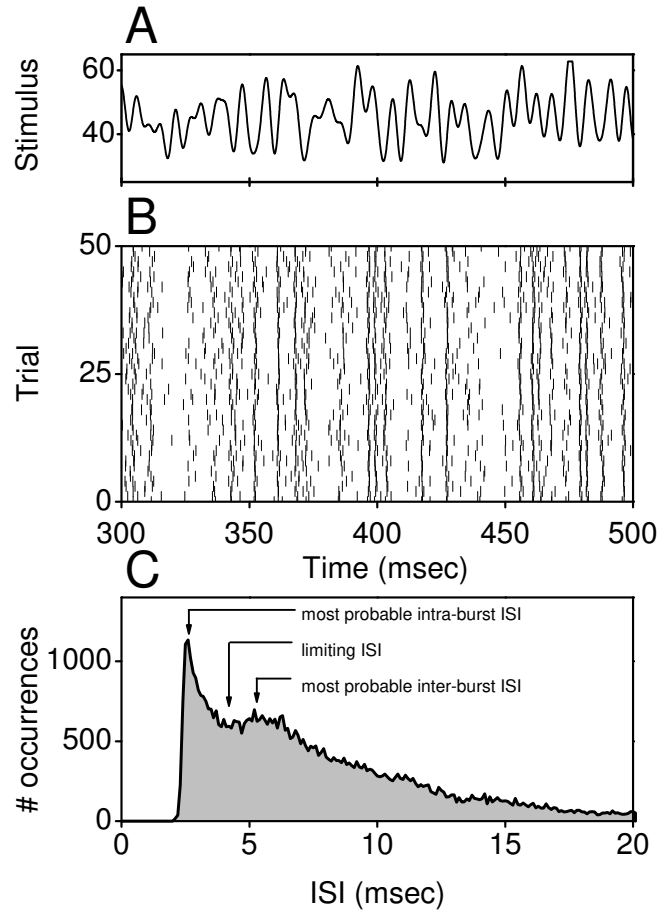}} \caption{\label{fig4} Recordings
in a grasshopper acoustic receptor. {\em A}: Amplitude of the sound
stimulus used to drive the recorded cell, in decibels. In the
experiment, each trial lasted for 1 second. {\em B}: Responses to
the first 50 stimulus presentations (out of a total of 503 trials).
Single spikes can be seen, as well as doublets and triplets. {\em
C}: ISI distribution of the recorded data. The two peaks correspond
to the most probable intra-burst ISI (left) and the inter-burst ISI
(right). The minimum between the two maxima is the limiting ISI
$\tau$ used for classifying spikes into bursts.}
\end{figure}

Grasshopper acoustic receptors fire in response to
amplitude-modulated broad band signals (see Machens et al., 2001;
Gollisch et al., 2002; Rokem et al., 2006; for further details). In
Fig.~\ref{fig4}{\em A}
we show the modulation of a sound-wave stimulus used to drive the
recorded cells. In {\em B}, 50 trials of the response of a single
receptor may be seen (see Sect.~\ref{s3p3} for the experimental
details).  A visual inspection of the raw data allows us to
identify isolated spikes, as well as short sequences of 2, 3 or 4
spikes coming in succession. These patterns are reliably
maintained along trials. In this cell, hence, it makes sense to
ask how much information is transmitted by the collection of all
spikes, and to compare it to the one encoded by the intra-burst
spike count $n$, as was done above with the simulated data.

In the simulations, all intra-burst ISIs lasted exactly the same
interval $\tau$. Hence, it sufficed to detect all sequences of
spikes separated by $\tau$ to identify bursts. In the experiment,
however, burst detection becomes less straightforward, since the
ISIs separating consecutive spikes form a continuum. We then need
a strategy to decide which spikes belong to the same burst, and
which ones correspond to different bursts. To that end, we take
the ISI distribution of the cell, depicted in Fig.~\ref{fig4}{\em
C}. The first and most prominent peak corresponds to the most
probable intra-burst ISI. The second peak is associated to the
characteristic time of inter-burst ISIs. Hence, we take the
minimum separating the two peaks as the limiting time $\tau = 3.9$
msec separating intra-burst and inter-bursts. We next go through
the spike train, and whenever we find two spikes that are
separated by an interval that is smaller than $\tau$, we assign
those two spikes to the same burst (a doublet). If yet another
spike follows closely (within an interval smaller than $\tau$),
then we add it to the previous ones, and form a triplet, and so
forth. The longest burst in the spike train contained 6 spikes (in
Sect.~\ref{s3p3}, we provide the relative frequency of bursts of
different $n$ values). With this procedure, we represented the
spike train as a sequence of integer values, as in the second
string shown in Fig.~\ref{fig1}.

\begin{figure}[htdf]
\centerline{\includegraphics{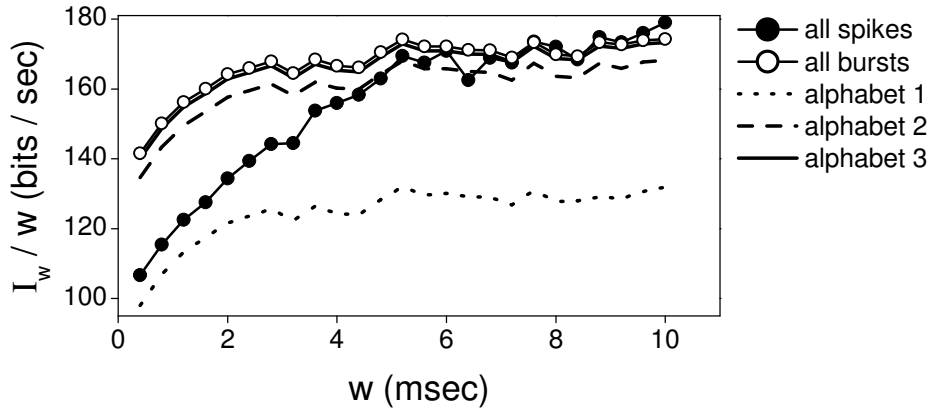}} \caption{\label{fig5}
Information ratio $I_w / w$ as a function of the word length $w$,
for the spike train generated by a grasshopper acoustic receptor.
Different lines and symbols correspond to different
identifications of the neural alphabet. The information
transmitted by all bursts is remarkably similar to that of all
spikes, implying that there is no information loss by reading out
the spike train in terms of a sequence of bursts.}
\end{figure}

In Fig.~\ref{fig5}, the ratio $I_w/w$ is depicted as a function of the
word length $w$. The black circles represent the information
transmitted by the whole collection of spikes. We see that $I_w/w$
starts at a fairly low value around 100 bits/sec, and
progressively grows up to almost 180 bits/sec, reaching the
asymptotic value at $w \approx 5$ msec. As explained above, $\tau$
was set to 3.9 msec. Hence, it proves crucial to work with windows
that are long enough to contain pairs of spikes, in order to
accomplish a correct readout of the neural message. Even longer
windows, incorporating triplets, quadruplets and higher-order
bursts, provide only minor corrections. The quantitative value of
such corrections is assessed by considering alternative codes. The
white circles in Fig.~\ref{fig5} correspond to the integer
$n$-based burst code, whereas the different lines represent the
reduced alphabets 1-3, introduced in Sect.~\ref{s4}. We see that
by completely ignoring the number of spikes within each burst
(alphabet 1) the transmitted information is 25\% lower. A
significant improvement is made by distinguishing single spikes
from the rest of the bursts (alphabet 2), and a minor refinement
is accomplished by further discriminating bursts of 2 spikes. The
overlap between the solid line (alphabet 3) and the white dots
(all bursts) implies that no further information is gained by
distinguishing between bursts of $n > 2$.

As dictated by the Data Processing Theorem (Cover and Thomas,
1991), when the word length $w$ is long enough to encompass all
information-bearing correlations in the spike train, the
information carried by the whole collection of spikes (black
circles) must lie above all the other codes, as confirmed by
Fig.~\ref{fig5}. However, in principle nothing prevents the code
based on all spikes to encode even more information than all the
burst-based codes. Indeed, if the precise location of subsequent
spikes inside each burst were relevant, the black circles should
be expected to lie significantly above all the other curves. From
Fig.~\ref{fig5}, however, we see that all the information in the
spike train is well captured by the burst-based code (white
circles). Hence, the integer representation of the spike train
carries all the relevant information. We therefore conclude that
the only response features that carry information are the location
of the first spike in each burst, and the number of spikes per
burst.

Having assigned a relevant value to the discrimination between
single spikes, doublets, and higher order bursts, we may now wonder
what these different symbols in the neural alphabet mean, in terms
of the stimulus features that elicit them. Going back to panels {\em
A} and {\em B}, we may now interpret the stimulus-response
transformation, by identifying the stimulus features that precede
single spikes, doublets, and higher-$n$ bursts. Specifically, we see
that doublets of spikes are generated by stimulus deflection of
higher amplitude than the ones required for single spikes. Triplets,
in turn, are preferentially found following stimulus excursions that
are not only high, but also wide. Quantitative methods to assess the
differences between the stimulus features eliciting bursts
containing different number of spikes have been described in
Eyherabide et al. (2008).

\section{Discussion}

Numerous studies have attempted to determine whether spike
patterns play a relevant role in the neural code. So far, two
basically different approaches have been employed. In the
so-called {\em quantitative} methods, the role of spike patterns
was assessed by means of information-theoretical measures. The
information transmitted by the full spike train is first
estimated, with as high a temporal resolution as limited-sampling
problems permit (see Panzeri et al., 2007, and references therein,
for a discussion of the sampling limitations of information
estimation). This full-blown information is regarded as the {\em
true} information content of the spike train. This information is
then compared to the one obtained by an altered readout mechanism
where spike patterns can no longer play a role. In some
experiments, the altered readout mechanism consists in considering
words containing only a single time bin (Reinagel and Reid, 2000;
Furukawa and Mindelbrooks, 2002; Kumbhani et al., 2007).
Obviously, single-bin words cannot account for the information
encoded in spike patterns. In other cases, the altered readout
mechanism is obtained by shuffling the spike train, mixing the
responses obtained throughout the different trials, within a fixed
time bin (Reinagel and Reid 2000, Furukawa and Middlebrooks, 2002;
Osborne et al., 2004; Montemurro et al., 2007). This procedure
eliminates within-trial structured spike patterns, though
preserving the correlations in the peri-stimulus time histogram.
If the amount of information obtained from the full spike train is
significantly higher than the one resulting from the altered
readout, it has often been concluded that spike patterns play a
relevant relevant role in information transmission.

The disadvantage of the quantitative approach is that, even if it
allows us to conclude that spike patterns are indeed important, it
does not enable us to identify which are the relevant code-words,
nor to disclose their meaning in terms of the stimulus. In this
respect, this method may be tagged as {\em blind}. In
compensation, being based on a numerical evaluation of the
information loss, one may precisely assess the relevance of the
spike patterns in quantitative terms. Therefore, although we still
lack explicit knowledge of the relevant code-words, we are
provided with a precise quantification of how much information is
lost, if those patterns (whichever they might be) are ignored.

As an alternative to the quantitative methods, there are also {\em
qualitative} ones. In the first place, we need to identify specific
patterns in the spike train, whose meaning we are interested in.
These patterns are picked either by visual inspection of the
responses, or with more sophisticated detection procedures (see for
example, 2001; Fellous et al., 2004; Eyherabide et al., 2008). Next,
one identifies the average stimulus feature eliciting the chosen
patterns, and performs some statistical test to assess whether these
features are different from one another or not (see, for example,
Reich et al., 2000; Arganda et al., 2007; Oswald et al., 2007;
Eyherabide et al., 2008). This procedure allows us to read out the
correspondence between specific patterns in the spike train and
their associated stimulus features. However, unless a quantitative
evaluation of the importance of each pattern is performed, one
cannot determine whether the correspondence between spike patterns
and stimulus features is a crucial ingredient of the code or not.
Moreover, if not all of the relevant response features have been
identified, such correspondence may remain incomplete, perhaps
missing even the most important code-words employed by the cell.

Here, we combined the quantitative and qualitative methods. We chose
specific spike patterns in the neural alphabet, thereby allowing an
explicit description of the neural code. Once these patterns have
been identified, a qualitative analysis of their meaning in terms of
the stimulus can easily follow, by means of covariance analysis. At
the same time, by constructing a sequence of nested alphabets, with
increasingly finer distinction between the patterns, we
quantitatively assessed the role of each pattern in the neural code.
Our goal was to discover the most compact code compatible with
preserving the available information.

In the example discussed in Sect.~\ref{s4}, we saw that there was
a representation of the spike train (the $n$-based alphabet) that
allowed us to recover {\em all} the information in the spike
train. If this alphabet was reduced, some information was lost. We
therefore conclude that the $n$-based alphabet contained all the
relevant patterns in the neural code, and no superfluous patterns.
When we applied the same procedure to the experimental analysis of
Sect.~\ref{s5} (where the relevant patterns were unknown) we found
that the relevant code-words were isolated spikes, doublets, and
bursts of $n > 2$. These three patterns contained all the
information in the spike train, and none of them was idle.

An important point in our procedure was to analyze the dependence of
$I_w/w$ on the window length $w$. Only when $I_w/w$ has reached its
asymptotic value can one be sure that the read-out windows are long
enough to encompass all information-bearing spike patterns. By
determining the minimal word length $w_{0}$ for which $I_w/w$ is
indistinguishable from its asymptotic value, we bound the maximal
length of the relevant patterns. In our work, conclusions about the
nature of the neural code are only drawn for $w \ge w_0$. In other
studies, however, the very dependence of $I_w/w$ on the window
length $w$ was used to assess the synergistic or redundant nature of
the code. Specifically, if for long $w$, the information $I_w/w$ was
larger than $I_1/1$, patterns were assigned a synergistic role (see,
for example, Brenner et al., 2000; Reinagel and Reid, 2000; Liu et
al., 2001; Kumbhani et al., 2007). In view of the results obtained
in Sect.~\ref{s4}, we believe that those discussions should be
handled cautiously, as explained below.

\subsection*{Consequences to the synergy/redundancy discussion}

\begin{figure}[htdf]
\centerline{\includegraphics{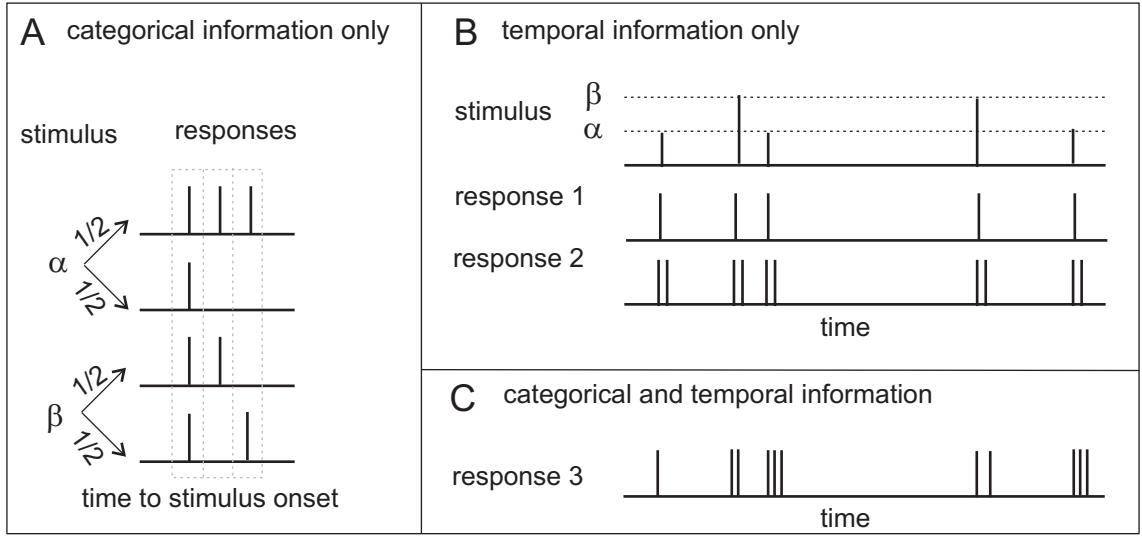}} \caption{\label{fig6}
Idealized experiments, to discuss the interplay between
categorical and temporal information. In {\em A}, each trial
contains only a single stimulus. Here, only categorical
information can be extracted from the responses. Individual time
bins contain no information at all, so the code is synergistic.
{\em B}: Numerous stimuli are shown in each trial, at random
times. Two possible noiseless codes are depicted. In both cases,
the responses identify the apparition of each stimulus, but they
do not encode the stimulus category. Response 1 allocates a single
spike to each stimulus, and response 2 a doublet. In response 1,
different time bins are neither synergistic nor redundant, whereas
in response 2, they are redundant. {\em C}: Example spike train
corresponding to the code of {\em A} and the stimulation protocol
in {\em B}.}
\end{figure}

In Sect.~\ref{s4}, readouts based on short words were shown to
confound categorical information with temporal information. In
order to disentangle these two aspects, here we consider three
more extreme experiments. The first one contains categorical
information alone, the second one, only temporal information, and
the third one, a mixture of the two. In Fig.~\ref{fig6}{\em A},
the response properties of the categorical experiment are
displayed. Each trial begins with the presentation of either
stimulus $\alpha$ or $\beta$. Responses are measured in a time
window locked to stimulus onset. In this case, hence, all response
properties encode stimulus identity, since in all trials, the
stimulus appears at time 0. Upon presentation of stimulus
$\alpha$, the cell either generates a triplet, or a single spike,
both options with probability 1/2. In response to stimulus
$\beta$, instead, one always gets a doublet, though the precise
timing of the second spike is uncertain. Using the time bins
depicted in dashed lines, different time bins are synergistic: the
information obtained from reading out the whole response is higher
than the sum of the information obtained in each time bin. In
fact, single bins contain no information at all.

Now let us consider an example where the neural responses encode
temporal information alone, as depicted in Fig.~\ref{fig6}{\em B}.
There, each trial contains many stimuli, and the time of stimulus
apparition is a stochastic variable itself. Two possible noiseless
codes are discussed. In both cases, responses are assumed to only
encode stimulus apparition, with no discrimination of stimulus
category. In Response 1, each stimulus is encoded with a single
spike, and in Response 2, with a doublet. Time bins are taken
small enough to allocate a single spike at most. If stimuli are
drawn independently from one another, in response 1 different time
bins are neither synergistic nor redundant. In response 2,
instead, different time bins are redundant. In general terms,
whenever a neural response encodes temporal information alone,
spike patterns are bound to introduce redundancy, since the
addition of extra spikes into the encoding patterns can add no new
information. Indeed, the larger the number of spikes in each
pattern, the larger the amount of redundancy. In {\em C}, we show
the response that would be obtained if the code introduced in {\em
A} was recorded with the time-dependent protocol of panel {\em B}.
In this case, although the encoding of categorical information is
synergistic, the time-dependent code is redundant. This is usually
the case, since for fine temporal binnings and low jitter,
categorical information is much smaller than temporal information
(see Eqs.~(\ref{edescompsition}) and (\ref{ecategorical}), where
the categorical information is shown to be proportional to the
apparition rate of the stimulus, which is usually much smaller
than unity). In conclusion, the tradeoff between the encoding of
stimulus category and the time of stimulus apparition is
responsible for the redundancy between different time bins, even
in neural codes where {\em all} the categorical information is
encoded in spike patterns.

\section{Conclusions}

We have shown that by studying how the information rate depends on
the choice of the neural alphabet we may decide which are the
important, information-bearing patterns in the neural code. It is
then possible to go back to the stimulus, and interpret those
code-words in terms of specific stimulus features. Our procedure,
hence, allows us to read-out the neural code, and at the same
time, to quantify the additional information encoded by specific
spike patterns. Our results are easily applicable to the case of
bursting sensory neurons, where the number of spikes inside each
burst has been reported to encode specific stimulus features.

We also discuss an example where, even though by construction all
the categorical information about the stimulus is encoded in spike
patterns, the code appears to be redundant when evaluated with
standard techniques. This follows from a tradeoff between the
synergy associated to the encoding of categorical information and
the redundancy in the representation of temporal information. We
therefore claim that the results obtained from estimating the
amount of synergy in the neural code need to be viewed as the
combined effect of these two opposing phenomena.

\section*{Acknowledgements}

We are deeply indebted to Ariel Rokem and Andreas V. M Herz, who
recorded the experimental data and contributed to this work with
valuable ideas and discussions. We also thank Marcelo Montemurro,
for his useful comments. Our research was supported by CONICET,
the ANPCyT, Universidad Nacional de Cuyo, and Alexander von
Humboldt Foundation.

\section*{References}

\begin{enumerate}

\item[-] Arabzadeh E., Panzeri S., and Diamond M. E. (2006)
Deciphering the spike Train of a Sensory Neuron: Counts and
Temporal Patterns in the Rat Whisker Pathway. J. Neurosci. 26,
9216-9226.

\item[-] Arganda S., Guantes R., and de Polavieja G. G. (2007).
Sodium pumps adapt spike bursting to stimulus statistics. Nature
Neuroscience 10, 1467-1473.

\item[-] Balakrishnan R., von Helversen D., von Helversen O.
(2001) Song pattern recognition in the grasshopper Chorthippus
biguttulus: the mechanism of syllable onset and offset detection,
J Comp Physiol [A]. 2001 May;187(4):255-64.

\item[-] Brenner N., Strong S. P., Koberle R., and Bialek W.
(2000) Synergy in the neural code. Neural Computation 12, 1531-1552.

\item[-] Cover T., and Thomas J. (1991) Elements of information
theory. Wiley and sons, New York.

\item[-] Denning K. S., and Reinagel P. (2005) Visual Control of
Burst Priming in the Anesthetized Lateral Geniculate Nucleus. J.
Neurosci. 25, 3531–3538.

\item[-] Eyherabide H. G., Rokem A., Herz A. V. M, and Samengo
I. (2008). Burst firing is a neural code in an insect auditory
system. Front. Comput. Neurosci. 2:3. \\
doi:10.3389/neuro.10.003.2008

\item[-] Fellous, J. M., Tiesinga P. H. E., Thomas P. J., and
Sejnowski T. J. (2004) Discovering Spike Patterns in Neuronal
Responses. J. Neurosci. 24, 2989-3001.

\item[-] Furukawa S., and Middlebrooks J. C. (2002) Cortical
Representation of Auditory Space: In\-for\-ma\-tion-Bearing
Features of Spike Patterns. J Neurophysiol 87, 1749-1762.

\item[-] Gawne T. J., and Richmond B. J. (1993) How independent
are the messages carried by adjacent inferior temporal cortical
neurons? J. Neurosci. 13, 2758-2771.

\item[-] Gollisch T., Sch\"utze H., Benda J., Herz A. V. M. (2002)
Energy Integration Describes Sound-Intensity Coding in an Insect
Auditory System. J. Neurosci 22: 10434-10448.

\item[-] Gollisch T., and Meister M. (2008) Rapid neural coding in
the retina with relative spike latencies. Science 319. 1108-1111.

\item[-] Kepecs A., and Lisman J. (2003). Information encoding and
computation with spikes and bursts. Network: Comput Neu Sys 14,
103-118.

\item[-] Kepecs A., Wang X. J., and Lisman J. (2002). Bursting
neurons signal input slope. J Neurosci 22, 9053-9062.

\item[-] Krahe R., Budinger E., Ronacher B. (2002) Coding of a
sexually dimorphic song feature by auditory interneurons of
grasshoppers: the role of leading inhibition. J Comp Physiol A
187: 977-985.

\item[-] Krahe R., Gabbiani F. (2004) Burst firing in sensory
systems. Nat Rev Neurosci 5:13-23.

\item[-] R. D. Kumbhani, Nolt M. J., and Palmer L. A. (2007)
Precision, Reliability, and Information-Theoretic Analysis of
Visual Thalamocortical Neurons. J. Neurophysiol. 98, 2647–2663.

\item[-] Lesica N. A., and Stanley G. B. (2004). Encoding of Natural
Scene Movies by Tonic and Burst Spikes in the Lateral Geniculate
Nucleus. J. Neurosci. 24, 10731-10740.

\item[-] Lesica N. A., Weng C., Jin J., Yeh C. I., Alonso J. M.,
and Stanley G. B. (2006). Dynamic Encoding of Natural Luminance
Sequences by LGN Bursts. PlosBiology 4 (7) e209.

\item[-] Liu R. C., Tzonev S., Rebrik S., and Miller K. D. (2001)
Variability and information in a neural code of the cat lateral
geniculate Nucleus. J. Neurophysiol. 86, 2789-2806.

\item[-] Machens C. K., Stemmler M. B., Prinz P., Krahe R.,
Ronacher B., and Herz A. V. M. (2001) Representation of Acoustic
Communication Signals by Insect Auditory Receptor Neurons. J.
Neurosci. 21: 3215–3227.

\item[-] Mainen, Z. F., and Sejnowski, T. J. (1996) Influence of
dendritic structure on firing pattern in model neocortical neurons.
Nature 382, 363-366.

\item[-] Metzner W., Koch C., Wessel R., and Gabbiani F. (1998).
Feature Extraction by Burst-Like Spike Patterns in Multiple Sensory
Maps. J. Neurosci. 18, 2283-2300.

\item[-] Montemurro M. A., Panzeri S., Maravall M., Alenda A.,
Bale M. R., Brambilla M., and Petersen R. S. (2007) Role of
precise spike timing in coding of dynamic vibrissa stimuli in
somatosensory thalamus. J Neurophysiol 98, 1871–1882.

\item[-] Nemenman I., Bialek W., de Ruyter van Steveninck R. R.
(2004) Entropy and information in neural spike trains: Progress on
the sampling problem. Physical Review E, 69: 056111.

\item[-] Osborne L. C., Bialek W., and Lisberger S. G. (2004) Time
Course of Information about Motion Direction in Visual Area MT of
Macaque Monkeys. J. Neurosci. 24, 3210 –3222.

\item[-] Oswald A. M. M., Chacron M. J., Dorion B., Bastian J.,
and Maler L. (2004) Parallel Processing of Sensory Input by Bursts
and Isolated Spikes. J. Neurosci. 24, 4351-4362.

\item[-] Oswald A. M. M., Doiron B., and Maler L. (2007) Interval
Coding. I. Burst Interspike Intervals as Indicators of Stimulus
Intensity. J Neurophysiol 97, 2731-2743.

\item[-] Panzeri S., Petersen R. S., Schultz S. R., Lebedev M.,
Diamond M. E. (2001) The role of spike timing in the coding of
stimulus location in rat somatosensory cortex. Neuron 29(3):
769-77.

\item[-] Panzeri S., Senatore R., Montemurro M. A., and Petersen
R. S. (2007) Correcting for the Sampling Bias Problem in Spike
Train Information Measures. J Neurophysiol 98: 1064–1072, 2007.

\item[-] Petersen R. S., Panzeri S., Diamond M. E. (2001)
Population Coding of Stimulus Location in Rat Somatosensory
Cortex. Neuron 32, 503-514.

\item[-] Pola G., Thiele A., Hoffmann K. P., Panzeri S. (2003)
An exact method to quantify the information transmitted by
different mechanisms of correlational coding.
Network 14, 35-60.

\item[-] Rokem A., Watzl S., Gollisch T., Stemmler M., Herz A. V.
M., and Samengo I. (2006) Spike-Timing Precision Underlies the
Coding Efficiency of Auditory Receptor Neurons. J. Neurophysiol.
95 2541-2552.

\item[-] Reich D. S., Mechler F., Purpura K.P., and Victor J. D.
(2000) Interspike intervals, receptive fields, and information
encoding in primary visual cortex. J. Neurosci. 20, 1964-1974.

\item[-] Reinagel P., Godwin D., Sherman S. M., Koch C. (1999).
Encoding of Visual Information by LGN Bursts. J. Neurophysiol. 81,
2558-2569.

\item[-] Reinagel P., and Reid R. C. (2000) Temporal coding of visual
information in the thalamus. J. Neurosci. 20, 5392-5400.

\item[-] Schneidman E., Bialek W., and Berry M. J. (2003) Synergy,
Redundancy, and Independence in Population Codes. J. Neurosci. 23,
11539-11553.

\item[-] Sherman S. M. (2001) Tonic and burst firing: dual modes of
thalamocortical relay. Trends in Neurosci. 24, 122-126.

\item[-] Strong S. P., Koberle R., de Ruyter van Steveninck R. R.,
and Bialek W. (1998) Entropy and information in neural spike trains.
Phys Rev Lett  80, 197-200.

\item[-] Theunissen F., and Miller J. P. (1995). Temporal encoding
in nervous systems: a rigorous definition. J. Comput. Neurosci. 2,
149-162.

\end{enumerate}

\end{document}